\documentstyle[psfig]{l-aa}

\begin{document}

   \thesaurus{07          % A&A Section 1: Letters to the 
                           editors
              (08.01.1;   % Stars: abundances
               08.05.3;   % Stars: evolution
               08.16.4;   % Stars: AGB and post-AGB
               08.09.2;   % Stars: individual: Sakurai's object
               08.09.2;   % Stars: individual: FG\,Sge
               08.22.3)}   % Stars: variables: R Coronae Borealis.

   \title{
A stellar endgame -- the born-again Sakurai's object}

   \author{M. Asplund\inst{1}, B. Gustafsson\inst{1}, 
D.L. Lambert\inst{2}, and N. Kameswara Rao\inst{3} }

   \offprints{M. Asplund (martin@astro.uu.se)}

   \institute{              
              Astronomiska observatoriet,
              Box 515,
              S--751 20 ~Uppsala,
              Sweden\\
              \and
              Department of Astronomy,
              University of Texas,
              Austin, TX 78712,
              USA \\
              \and
              Indian Institute of Astrophysics,
              Bangalore 560 034,
              India \\
              }

\date{Received: Feb. 7; accepted: March 13, 1997}

   \maketitle

   \markboth{Asplund et al.: 
A stellar endgame -- the born-again Sakurai's object}
{Asplund et al.: 
A stellar endgame -- the born-again Sakurai's object}

\begin{abstract}                                                                           
  
The surface chemical composition of this remarkable star
shows that it is  hydrogen-deficient,
carbon-rich and enriched in the light $s$-process elements.
Spectra taken in May and October 1996 indicate
a  decrease in the surface hydrogen abundance 
by 0.7\,dex in five months
along with an increase in the
abundances of Li, Sr, Y and Zr.
The abundance changes are in agreement with the hypothesis
of the star being a 
rapidly evolving ``born-again'' AGB star  
experiencing  a final He-shell flash, similar to FG\,Sge. 
The $^{12}$C/$^{13}$C ratio in October is very low,
also suggesting hydrogen ingestion.
By chemical composition, Sakurai's object resembles
the R Coronae Borealis (R\,CrB) stars. 
    

\end{abstract}

\section{Introduction}

The nova-like brightening of Sakurai's object in Sagittarius  has been 
attributed to 
the  final helium-shell flash of a central star of a planetary
nebula, that returns the star towards the domain of red giants in the
Hertzsprung-Russell diagram
(Duerbeck \& Benetti 1996).
Few  stars have been identified with
this phase of evolution:
examples include
FG Sge, V605 Aql, the planetary nebulae Abell 30, Abell 78 and N66. 
It is expected that a 
born-again red giant will consume hydrogen and become starkly
hydrogen-deficient, helium- and carbon-rich.
Low-resolution spectra led Duerbeck and Benetti (1996) to
suggest that Sakurai's object is hydrogen-poor. 
The presence of strong lines
of neutral carbon and oxygen was also noted.
 
Changes of the surface chemical composition may be rapid for born-again
AGB stars, as was observed for FG Sge. Sakurai's object offers the
prospect of monitoring such secular changes in another born-again candidate.
Observations, as reported here, are surely crucial for an improved understanding of the
final He-shell flash.

\section{Observations}

Spectra covering 3700-10150\,\AA $ $ at a resolution of about 30,000
were obtained with 
the 2.7\,m telescope at McDonald Observatory on May 5 and 6 
and on October 7,  1996. 
A spectrum was also obtained with the 2.1\,m telescope:
this spectrum from May 9, 1996, covers the region
5720\,\AA\ to 7200\,\AA\ at a resolution of about 60,000.

\section{Chemical composition}

Our analysis is based on 
line-blanketed, hydrogen-deficient
model atmospheres, similar to those
described by Asplund et al. (1997) but with a range of hydrogen
abundances.
In estimating the stellar parameters T$_{\rm eff}$, log $g$ and hydrogen
abundance  various ionization 
(Fe\,{\sc i}/Fe\,{\sc ii}, Mg\,{\sc i}/Mg\,{\sc ii}, 
Si\,{\sc i}/Si\,{\sc ii},
Cr\,{\sc i}/Cr\,{\sc ii}) 
and excitation equilibria 
([O\,{\sc i}]/O\,{\sc i}, Fe\,{\sc i}, Fe\,{\sc ii})
together with the H$\beta$ and H$\alpha$ 
line profiles (with line broadening data following Seaton 1990) 
have been used.
The C/He ratio was determined from the C\,{\sc ii} and He\,{\sc i} lines
in the May spectra, which indicate C/He $\simeq 10$ \%.
The same ratio had to be assumed for October when 
the 
lines were too weak to be utilized.
The microturbulence parameter 
was estimated from Ti\,{\sc ii}, Fe\,{\sc i} and Fe\,{\sc ii}
lines of different strengths.
The May spectra are characterized by $T_{\rm eff} = 7500\pm 300$\,K, 
log\,$g = 0.0\pm 0.3$, and
$\xi_{\rm t} = 8.0\pm 1.0$\,km\,s$^{-1}$, while it had cooled significantly 
in October: $T_{\rm eff} = 6900$\,K, log\,$g = 0.5$, and
$\xi_{\rm t} = 6.5$\,km\,s$^{-1}$. 
In fact, the derived parameters are not consistent with a constant stellar
luminosity but rather indicate a decrease by a factor of 4,
which is not supported by the observed photometry.
It could, however, 
be that hydrostatic equilibrium is  
inapplicable in May due to an expansion of the star or effects of
turbulent pressure: 
a dynamical atmosphere can be mimicked by an underestimate of log\,$g$ 
when assuming hydrostatic equilibrium.
Indeed, with the May parameters the star is located at the 
classical Eddington limit 
(e.g. Asplund \& Gustafsson 1996).

\begin{figure}[t]
\centerline{
\psfig{figure=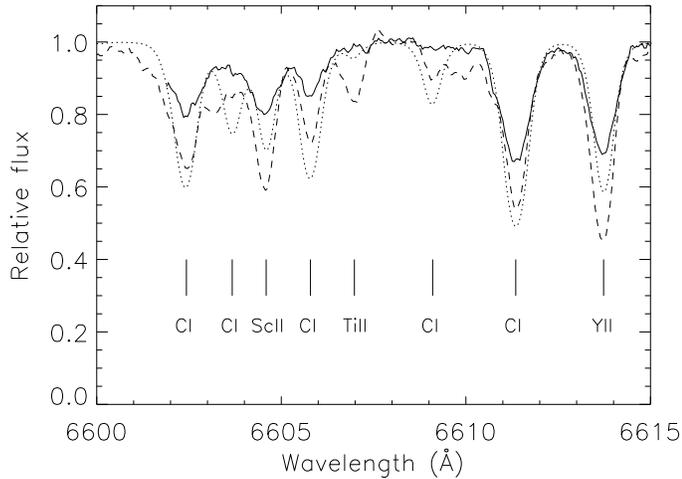,height=6.cm}}
\caption{A selected piece of spectrum in May (solid) and October (dashed)
showing the increase of some elements, e.g. Sc, Ti, and Y. The dotted
curve is the synthetic spectrum with the stellar parameters of 
October but with the May abundances.
Note also that all predicted C\,{\sc i} lines are too strong
}
         \label{f:spectra}
\end{figure}

\begin{figure}[t]
\centerline{
\psfig{figure=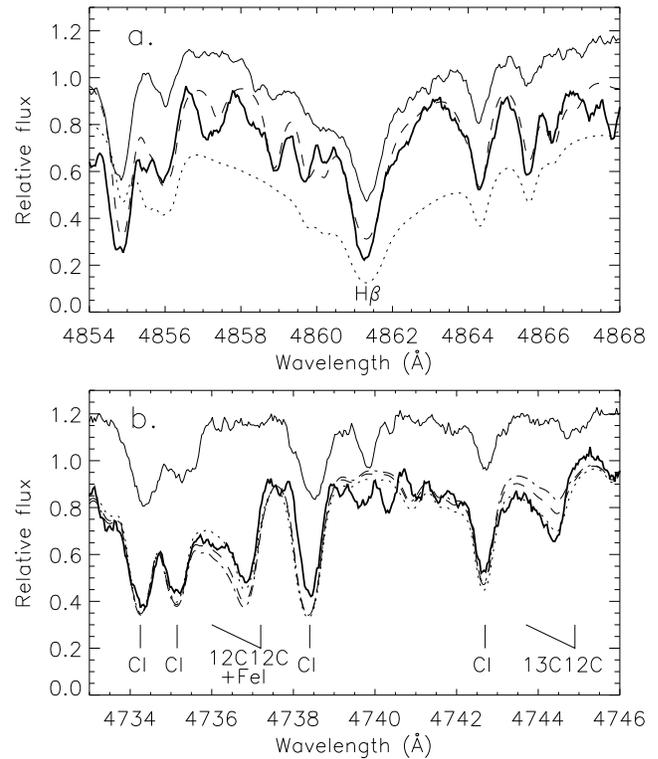,height=9.8cm}}
\caption{{\bf a} H$\beta$ in October (thick solid) 
compared with predicted
line profiles for solar H (dotted) and H-deficient by 
3.0\,dex (dashed).
{\bf b} C$_2$ (1-0) Swan band for $^{12}$C/$^{13}$C = 2 (dotted),
5 (dashed) and 10 (dash-dotted), together with the 
observed October spectrum
(thick solid). 
Also shown in both figures are the May spectra (solid) but 
displaced upwards by 0.2 for clarity
}
         \label{f:hbeta}
\end{figure}

The analysis of the C\,{\sc i} lines reveals the same inconsistency between
theoretical and observed line strengths as for R\,CrB stars (Gustafsson \& Asplund 1996;
Lambert et al., in preparation): the strengths of weak lines predicted
with the input C abundance are a factor of 4 stronger than observed
(Fig. \ref{f:spectra} and \ref{f:hbeta}). 
It should be noted that no agreement 
between all $T_{\rm eff}$-log\,$g$ indicators
could be achieved using consistent C abundance for the analysis.
Naturally, this C\,{\sc i} problem makes the absolute abundances uncertain
but relative abundances
are generally expected to be much less affected (Lambert et al., in preparation).

The derived LTE
abundances for May and October are summarized in Table \ref{t:abund}.
More details on the analysis and atomic data (lines, gf, hfs, etc) as
well as a comparison with V854\,Cen
will be given elsewhere.
The weak Balmer lines certainly 
rule out a solar hydrogen abundance (Fig. \ref{f:hbeta}).
Note that the absolute abundances of most elements 
are effectively unchanged 
from May to October within the uncertainties (typically $\leq 0.3\,$dex).
Some elements, however, exhibit a marked change, for example, hydrogen declined
as lithium and the light $s$-process elements increased
in abundance 
by a factor of about 4 (Fig. \ref{f:spectra}).
Also Sc, Ti, Cr and Zn seem to have increased during the time\-span
(Fig. \ref{f:spectra}). 
The general agreement between the May and October abundances for most
elements suggests that the stellar parameters are not seriously in
error, which could 
otherwise 
have resulted in spurious abundance
effects. 
Besides Li, the abundances of 
elements showing variations are not very sensitive
to the stellar parameters:
the required $\Delta T_{\rm eff} \approx 1000$\,K for
either May or October to annul the abundance variations 
would be inconsistent with the $T_{\rm eff}$--log\,$g$ indicators
and introduce other as severe changes (e.g. for Ca)
less easily explainable;
a different log\,$g$ can not simultaneously explain 
all changes.
It would also only aggravate the luminosity discrepancy.
Hence, the few 
changes seem to be real. Furthermore, they are 
limited to elements expected to show alterations due
to a final flash.

\begin{table}[t]
\caption{ Chemical compositions of Sakurai's object, the 
R\,CrB stars and the Sun
(normalized to log\,($\Sigma \mu_i \epsilon_i$) = 12.15)
\label{t:abund}
}
\begin{tabular}{lccccc} 
 \hline
  Element & Sun$^{\rm a}$ & \multicolumn {2} {c} {Sakurai's object} & 
\multicolumn {2} {c} {R\,CrB$^{\rm b}$}   \\ 
% \cline{6-9} \\
 && May & October & majority & minority  \\             
\hline \\
H  & 12.0 &  9.7 & 9.0 &   &   $<4.1 - 10.8$   \\
He & 11.0 & 11.4$^{\rm c}$ & 11.4$^{\rm c}$ & 11.5$^{\rm c}$ & 11.5$^{\rm c}$   \\
Li &  3.3$^{\rm a}$ &  3.6 & 4.2 &       &   \\ 
C  &  8.6 &  9.7$^{\rm d}$ & 9.8$^{\rm d}$ & 8.9$^{\rm d}$ & 8.6 -- 9.5$^{\rm d}$ \\

N  &  8.0 &  8.9 &  8.9 &  8.6 &  7.6 -- 8.6 \\  
O  &  8.9 &  9.5 &  9.4 &  8.2 &  7.5 -- 8.8 \\
Ne &  8.1 &  9.3 &      &      &  7.9 -- 9.6 \\
Na &  6.3 &  6.7 &  6.8 &  6.1 &  5.8 -- 5.9 \\
Mg &  7.6 &  6.6 &  6.5 &      &  6.1 -- 7.3 \\
Al &  6.5 &  6.6 &  6.3 &  6.0 &  5.3 -- 5.6 \\  
Si &  7.5 &  7.1 &  7.5 &  7.1 &  7.3 -- 8.1 \\ 
S  &  7.3 &  6.6 &  6.9 &  6.9 &  6.7 -- 7.6 \\          
K  &  5.1 &  4.8 &  5.0 &      &            \\
Ca &  6.4 &  5.6 &  5.5 &  5.4 &  5.0 -- 5.3 \\  
Sc &  3.2 &  3.1 &  3.9 &      &            \\
Ti &  5.0 &  4.1 &  4.6 &      &            \\
Cr &  5.7 &  4.5 &  5.1 &      &            \\
Fe &  7.5 &  6.3 &  6.6 &  6.5 &  5.0 -- 5.8 \\  
Ni &  6.2 &  6.1 &  6.2 &  5.9 &  5.2 -- 5.8 \\  
Cu &  4.2 &  4.9 &  5.0 &      &            \\
Zn &  4.6 &  4.7 &  5.4 &  4.3 &  3.8 -- 4.1 \\ 
Rb &  2.6 & $<3.7$ &  4.6 &      &            \\
Sr &  3.0 &  4.9 &  5.4: &      &            \\
Y  &  2.2 &  3.3 &  4.2 &  2.1 &  0.6 -- 2.8 \\            
Zr &  2.6 &  3.0 &  3.5 &      &            \\
Ba &  2.1 &  1.5 &  1.9 &  1.6 &  0.7 -- 1.3 \\  
La &  1.2 & $<1.6$ &  1.5 &      &  \\
\hline
\end{tabular}

\begin{list}{}{}
\item[$^{\rm a}$] From Grevesse et al. (1996). For Li the meteoritic 
value is adopted.
\item[$^{\rm b}$] From Rao \& Lambert (1996) and Jeffery \& Heber (1993). 
The majority is an average of 14 stars while 
the minority consists of V\,CrA, VZ\,Sgr, V3795\,Sgr and DY\,Cen.
\item[$^{\rm c}$] Input C/He ratio for model atmospheres: C/He=1\% assumed for 
R\,CrB stars and 10\% estimated for Sakurai's object from the 1996 May spectra.
\item[$^{\rm d}$] Spectroscopically determined C\,{\sc i} 
abundance, see text.
\end{list}

\end{table}

The metallicity of Sakurai's object is, judging from Fe, slightly below
solar by 0.2\,dex in mass fraction (0.9\,dex if the input rather than
the spectroscopic C abundance is adopted).
The quantities [Si/Fe], [S/Fe], [Ca/Fe],
and [Ti/Fe], which are 0.8, 0.6, 0.3 and 0.4 
respectively, are, if unchanged
from the star's birth, also 
indicative of a metal-poor star (Edvardsson et al. 1993).
An isotopic ratio $1.5\leq^{12}$C/$^{13}$C$ \leq 5$ is determined from
the strong C$_2$ (1-0) and (0-1) Swan bands (Fig. \ref{f:hbeta}).
The strengthening of the C$_2$ bands due to the change
in stellar parameters is clearly illustrated in Fig. \ref{f:hbeta}.

It is of considerable interest to compare the compositions of Sakurai's object
and FG\,Sge, another born-again candidate which has 
recently experienced R\,CrB-like visual declines. FG\,Sge resembles Sakurai's
object in that it is
strongly $s$-element enriched (Langer et al. 1974), as well as
carbon-rich and poor in iron-group elements, except for Sc
(Kipper \& Kipper 1993).  In FG Sge, however,  
the heavy $s$-elements are as
overabundant as the light, and it has not yet been shown to be
hydrogen-deficient. 
FG\,Sge may therefore have experienced a late shell flash as a luminous post-AGB
star rather than a final flash as a white dwarf
(Bl\"ocker \& Sch\"onberner 1996).
  
Two of the outstanding aspects of the chemical composition of Sakurai's
object are hallmarks of the R\,CrBs: H-deficiency and a high C content,
but also other similarities in relative abundances exist
(Lambert et al., in preparation; Rao \& Lambert 1996; 
Lambert \& Rao 1994).
Except for the high Y/Fe other observed X/Fe ratios are similar
to those found in R\,CrB stars.
In particular it resembles the (relatively) H-rich V854\,Cen
(Asplund et al., in preparation).
If, however, C/He is correctly estimated, it may sooner be related 
to objects such as V605\,Aql, Abell 30 and 78 and the hot
R\,CrB star V348\,Sgr, which are also surrounded by planetary
nebulae and have been proposed to be final flash candidates
(Renzini 1990).

Similar abundance patterns as presented here for Sakurai's object
have also been obtained in less detailed analyses by
Shetrone \& Keane (1997) and Kipper \& Klochkova (1997). 
Shetrone \& Keane's finding of a 
near normal H abundance is however very puzzling.

\section{Abundance variations and nucleosynthesis}
                                                                                                 
In broad terms, the composition of Sakurai's object shows evidence of
severe contamination by material exposed to hydrogen and helium burning
and associated nuclear reactions. Close examination provides 
some interesting
constraints on the nucleosynthesis experienced by the star.

The present atmosphere is not
a simple mix of initial unprocessed gas, gas
run through the H-burning CNO-cycles, and  H-exhausted gas exposed to He-burning,
but must have been accompanied by further processing.
This is
demonstrated by the low observed $^{12}$C/$^{13}$C ratio, which
encompasses the equilibrium value of 3.5 for CNO-cycling.
As the equilibrium abundance of $^{13}$C is very low following He-burning, 
the observed ratio suggests that $^{12}$C from He-burning has been
exposed to hot protons. It would seem that
C-rich material from He-burning has been mixed with 
ingested hydrogen such that
the proton supply is effectively exhausted in converting 
inhibited (see Renzini 1990).

Not all protons are consumed in He-rich regions.
Production of lithium is ascribable to the
Cameron-Fowler (1971) mechanism. 
Here  $^3$He synthesised in a low mass main sequence star is
converted to $^7$Li
in an  envelope that convects $^7$Li to low temperatures 
where it survives
until re-exposed to high temperatures. 
Production of lithium implies
H-burning in regions not previously exposed to 
H-burning temperatures; $^3$He
which is destroyed in regions that have undergone H-shell or H-core burning 
can hardly be resynthesised.
The observed Li is not a fossil from an 
earlier stage as a  Li-rich AGB star: 
the predicted Li/H ratio for 
AGB stars 
which have undergone hot-bottom burning
is 10$^{-8}$ while the
observed ratio 
is 10$^{-5}$ to 10$^{-6}$ and hydrogen consumption
necessarily destroys fossil lithium.
The overabundant Na and Al have likely been synthesised through 
$^{22}$Ne(p,$\gamma$)$^{23}$Na and $^{25}$Mg(p,$\gamma$)$^{26}$Al.
As in other H-deficient stars Ne is very high,
which can not be explained by $\alpha$-captures on N from initial
CNO, but must be due to products of He-burning and possibly 
additional CNO-cycling.

A remarkable feature of Sakurai's object is the large overabundance of
light $s$-process elements and the high ratio of light to heavy $s$-process
elements. Probably, $^{13}$C($\alpha,n)^{16}$O is the neutron source. 
The $s$-processing may be characterized by the neutron exposure $\tau$.
We find a good fit to the 
abundances from Ni to La for October with $\tau =  0.2 \pm 0.1$\,mb$^{-1}$ 
using Malaney's (1987) predictions for a single exposure.
The Rb abundance indicates a low neutron density of 
$N_{\rm n} \approx 10^8$\,cm$^{-3}$ (Malaney 1987), while 
no useful limit could be set on the Tc abundance.
An exponential distribution of exposures provides less
good agreement with the observed abundances. 
For the R\,CrB star
U Aqr, which also shows light $s$-element enhancements,
Bond et al. (1979) obtained $\tau \simeq 0.6$\,mb$^{-1}$. Such exposures
imply that about 10 neutrons were captured by each Fe seed nucleus.
Given that the observed ratio $^{13}$C/Fe $\simeq 10^3$, the exposure, even
in the presence of neutron poisons such as $^{14}$N, seems an achievable
goal. The fact that Ni, Cu and Zn are well fit by the predictions indicates
that the envelope consists mostly of material exposed to neutrons. This fact also
likely explains the anomalous high ratios of K/Fe and Sc/Fe.

The final He-shell flash may occur in a luminous post-AGB star or the
white dwarf that evolves from the post-AGB star.
In the latter case, hydrogen may be mixed  with deep layers of He 
and C and consumed. In contrast, the H-burning layer in the post-AGB star
prevents deep mixing. About 10\% of all AGB stars may experience their final
He-shell flash as a white dwarf and, if H consumption is severe, may
convert the born-again AGB star to an R\,CrB star 
(Renzini 1990; Iben et al. 1996).
Iben \& MacDonald (1995) have presented a model in which mixing and
nucleosynthesis were followed: their chosen model of a 0.6\,$M_{\odot}$ star
ended with an outer layer having the abundance ratios (by number of atoms)
H/He $\simeq 10^{-0.8}$,
C/He $\simeq 10^{-1.2}$,
N/C $\simeq 10^{-0.5}$, and O/C $\simeq 10^{-1.3}$. 
This resembles the
composition of Sakurai's object, 
apart from the predicted H deficiency 
not being as severe as observed.
The model O/C ratio is lower than observed but might
be raised by adjustment of the uncertain rate for the reaction
$^{12}$C($\alpha,\gamma$)$^{16}$O. 
In summary, final flash models offer a 
tantalising 
prospect of accounting for 
the observed composition of Sakurai's object.

Life as a born-again AGB star is brief: the model by Iben \& MacDonald
(1995) brightens by a factor of 10 and cools from T$_{\rm eff}$ of 40,000K
to 6300K in just 17 yr. Evolution over the narrow temperature range
covered by Sakurai's object between May and October is, of course, much
faster. 
The evolutionary timescale seems similar to that of V605\,Aql
(Lundmark 1921).
The  timescale for compositional changes for Sakurai's object is
likely even shorter; processed material rising from below will
mix very quickly with the atmosphere.
The other 
final flash candidate FG\,Sge has also
showed rapid abundance alterations, e.g. some $s$-process elements
increased by 0.8 dex in 7 years (Langer et al. 1974).

It is now important  to extend the few available calculations
of the final flash to a wider range of 
initial conditions, and to include Li-production and $s$-processing. The hints
that the surface composition is evolving rapidly must be pursued by
continued spectroscopic observations, 
which may shed further light on its evolutionary 
status and relation to the R\,CrB stars;
we may have witnessed the birth of an R\,CrB star. 
Furthermore, a determination of the nebular composition would reveal the original
composition of Sakurai's object prior to the final flash.
Monitoring the visual variability of the star 
searching for R\,CrB-like declines is naturally of
importance.

\begin{acknowledgements}
We are grateful to Craig Wheeler for bringing the star to our attention, to
Guillermo Gonzalez for a spectrum of Sakurai's object,
and to Jim MacDonald and Icko Iben for helpful comments. 
Nikolai Piskunov and Sveneric Johansson are thanked for help with the
hydrogen profiles and atomic data, and 
V.P. Arkhipova and V.P. Goranskij for photometric information.
Financial support from the Swedish
Natural Research Council, NSF (grant
AST93-15124) and the Robert A. Welch Foundation of Houston, Texas
is acknowledged. 
The analysis has made use of the VALD database.
The referee Simon Jeffery is thanked for helpful comments.
\end{acknowledgements}

\end{document}